# A reconfigurable multiple-format coherent-dual-band signal generator based on a single optoelectronic oscillation cavity

Yibei Wang[1,2,3], Yalan Wang[1,3], Hongyi Wang[1], Xiaotong Liu[1], Hong Chen[1], Jin Zhang[1], Dongyu Li[1], Dangwei Wang[1], and Anle Wang[1,*]

[1]Microwave Photonics Center, Early Warning Academy, Wuhan 430017, China
[2]Hubei Key Laboratory of Ferroelectric and Dielectric Materials and Devices, Faculty of Physics and Electronic Science, Hubei University, Wuhan, 430062, China

[3]These authors contributed equally to this work.
*Corresponding author: anlehit@163.com

**Abstract:** An optoelectronic oscillation method with reconfigurable multiple formats for simultaneous generation of coherent dual-band signals is proposed and experimentally demonstrated. By introducing a compatible filtering mechanism based on stimulated Brillouin scattering (SBS) effect into a typical Phase-shifted grating Bragg fiber (PS-FBG) notch filtering cavity, dual mode-selection mechanisms which have independent frequency and time tuning mechanism can be constructed. By regulating three controllers, the proposed scheme can work in different states, named mode 1, mode 2 and mode 3. At mode 1 state, a dual single-frequency hopping signals is achieved with 50 ns hopping speed and flexible central frequency and pulse duration ratio. The mode 2 state is realized by applying the Fourier domain mode-locked (FDML) technology into the proposed optoelectrical oscillator, in which dual coherent pulsed single-frequency signal and broadband signal is generated simultaneously. The adjustability of the time duration of the single-frequency signal and the bandwidth of the broadband signal are shown and discussed. The mode 3 state is a dual broadband signal generator which is realized by injecting a triangular wave into the signal laser. The detection performance of the generated broadband signals has also been evaluated by the pulse compression and the phase noise figure. The proposed method may provide a multifunctional radar system signal generator based on the simply external controllers, which can realize low-phase-noise or multifunctional detection with high resolution imaging ability, especially in a complex interference environment.

## Introduction

In today's radar system, to achieve high-precision multi-dimensional measurement of targets at complex interference environment, a high-performance radar signal generator with flexible multiple-format reconfiguration is a basic necessity [1-4]. Employing broadband signal, for example, is a typical solution to realize high-precision imaging of the target combined with pulse compression method [5]. Pulsed single frequency signal is committed to be a typical solution to detect large range target with fast speed [6]. And the fast frequency-hopping signal can face the complex interference environment [7,8]. In a conventional electrical radar system, generation of complex reconfigurable microwave signal is urgent, while facing the electrical bottleneck in generating low phase noise signal or wide frequency tuning signal [9.10].

Optoelectronic oscillator (OEO) exudes an inherent advantage in generating low phase noise signals [11-15] since the first prototype exposed by S. Yao and L. Maleki [16]. Especially, recently novel optoelectronic oscillators are emerged to expand the generated waveform styles and performance for various application scenario [17-26]. Fourier domain mode-locked OEO (FDML-OEO) that directly generates a broadband signal from the OEO cavity breaks the mode construction time [27], which is discovered as a potential application method for wideband Radar system. On another aspect, frequency-hopping signal with fast hopping speed generated by a soliton OEO is brought up to adapt to the complex electromagnetic interference environments [28].

In this paper, a method with reconfigurable multiple formats for simultaneous generation of coherent dual band signal based on a simple OEO cavity is proposed and experimentally. A dual band MPF is constructed by combining the SBS effect and phase-shift grating, which confirms the flexible tunability of the scheme. By adopting three controllers, the simple OEO can work at three states without any changes of the structure. The MPF and proposed three mode working principles are discussed. At mode 1 state, a dual single frequency-hopping signal is achieved with 50 ns hopping speed. The central frequency flexibility is realized by the laser external current drivers. Especially, by adopting two polarization controllers (PCs) to regulate the power contributions of the dual band, the time duration of the higher frequency to that of the lower frequency is tuned form 1.56 μs to 3.38 μs. The mode 2 state is realized by applying the Fourier domain mode-locked technology into the proposed optoelectrical oscillator, in which dual coherent pulsed single frequency signal and broadband signal is achieved. The duration of single frequency signal is tuned from 0.78 μs to 1.43 μs by regulating the PCs. The bandwidth adjustability is achieved by changing the LD's external current driving voltage. Furthermore, the mode 3 state is a dual broadband signal generator which is realized by injecting a triangular wave into the signal laser. And by changing the wavelength of the signal light, the bandwidth and center frequency of the generated broadband signal can also be regulated. The proposed simple OEO can realize switch among different working states by just tune the external controllers without any changing of the structure. In addition, it is integratable and has great potential application in multifunction radar.

**Principle and operation**

The schematic of the proposed flexible multiform waveform generator is demonstrated in Fig. 1(a). The continuous wave (CW) light emitted by a laser diode (LD1) is sent into a phase modulator (PM). The modulated output optical signal passes through an optical isolator (ISO) and meets the pump optical signal emitted by the laser diode (LD2) in a highly nonlinear fiber (HNLF) with length of 1 Km. There are two polarization controllers (PCs) in the loop to control the polarization state of both the pump light signal and the light signal reflected by the notch based on PS-FBG, respectively. The filtered optical sidebands are demodulated at a photodetector (PD). Two EAs are cascaded to compensate the loop energy loss and a bandpass filter (BPF) is adopted to filter out the unwanted spurs. The generated microwave signal is then divided into two parts via an electrical coupler (EC). One part is fed back to the PM to form a closed loop, the other is connected to an analytical instrument to capture the frequency and time characteristics of the generated signal, such as Electric Signal Analyzer (ESA) and a digital storage Oscilloscope (OSC). There are three controllers in this setup to regulate the working state of the system. Controller 1 and controller 2 are the external current drivers of the LD1 and LD2, respectively. The controller 3 consisted of two PCs in the loop is to regulate the energy distribution. Fig. 1(b) is the illustration of the proposed three working modes operation principle. The upper picture illustrates the filter structure employed in the proposed scheme, in which the red line presents the reflection band of the PS-FBG. The $f_{signal}$, $f_{pump}$ and $f_{notch}$ is the lasing frequency of LD1, LD2 and the position of the notch of the PS-FBG, respectively. The lower picture is the illustration of the proposed three working states. At the mode 1 working state, a dual single-frequency MPF is constructed by converting the phase modulation into the intensity modulation through SBS effect and PS-FBG. The central frequencies of the dual-single MPF can be expressed as follows.

$$f_{MPF1} = f_{notch} - f_{signal}$$
$$f_{MPF2} = f_{gain} - f_{signal} \qquad (2)$$

where $f_{gain}$ is the gain spectrum position generated by SBS effect, $f_{MPF1}$ and $f_{MPF2}$ is central frequency of the dual-single MPF respectively. Moreover, two PCs (controller 3) are adopted to regulate the power contribution of the dual frequency, and the frequency hopping phenomenon can be achieved at proper condition. Since the frequency hopping is achieved by polarization state, the time duration ratio can be easily tuned by just controlling the two PCs in the loop. At the mode 2 working state, a scanning triangular wave (controller 1) is injected into the LD2 to control the pump light wavelength, obtaining a frequency-scanning MPF. Therefore, the mode 2 can achieve the generation of coherent signal with different frequency modulation characteristics, which is one single frequency signal and one broadband frequency signal. The central frequency of the single-frequency signal can be expressed as $f_{MPF2}$, while the central frequency of the broadband signal is sweeping repeatedly from $f_{MPF1}$ to $f'_{MPF1}$, as shown in mode 2 of Fig.1(b). And the bandwidth of the broadband signal can be regulated independently. The final mode 3 is achieved by scanning the signal lasing frequency of the LD1 through controller 2, which results in a coherent dual-broadband frequency signal. In this case, a scanning signal laser beat with the notch frequency of the PS-FBG, resulting in the first broadband signal. At the same time, the scanning laser is working as the signal laser to accomplish the SBS effect, achieving the second wideband signal. The central frequency of the broadband signal based on PS-FBG is sweeping repeatedly from $f_{MPF2}$ to $f'_{MPF2}$, while the central frequency of the broadband signal based on SBS is sweeping repeatedly from $f_{MPF1}$ to $f'_{MPF1}$. Since the two broadband signals come from one cavity, they are coherent and have the same characteristics. In a word, by just control the three external controllers without any changing of the system, a flexible multiform waveform generator is constructed.

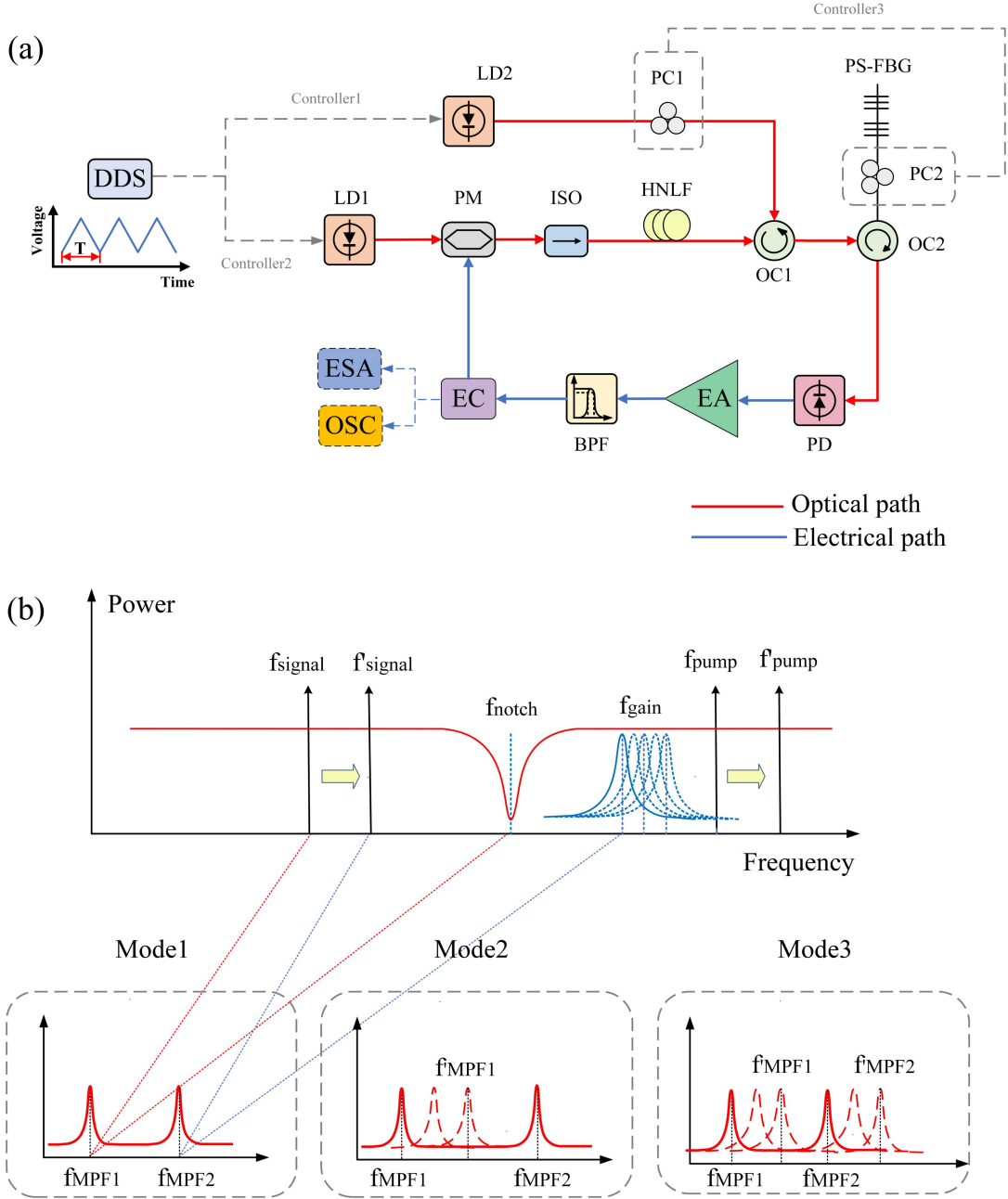

Figure 1 (a) The schematic of proposed method for simultaneous generation of coherent dual band signal. LD, laser diode; PM, phase modulator; ISO, isolator; PC, polarization controller; OC, optical coupler; PD, photodetector; EA, electrical amplifier; BPF, electrical bandpass filter; EC, electrical coupler; OSC, oscilloscope; ESA, electrical spectrum analyzer; (b) Operation principle of the proposed three working modes.

**Experiments and results**

A LD with maximum output power of 16 dBm (ID Koheras BASIK E15) and a multi-channel tunable LD with output power of 12dBm (ID Thorlabs CLD1015) are used as the probe light source and pump light source, respectively. The probe light is modulated in a PM (MPZ-LN-40, iXblue Photonics) with a bandwidth of 40 GHz. The modulated light then passes through a length of 1 km HNLF as the probe signal. The pump light is launched into the HNLF via an OC to generate a Brillouin gain region. Then the +1 sideband of the modulated signal falls into the

Brillouin gain region and the notch of PS-FBG centered at 1550.094 nm to accomplish the conversion from phase modulation to intensity modulation. An isolator is inserted after PM to protect the probe signal laser. A tunable ATT (DVOA-1000, OVLINK) is used to control the optical power to ensure stable oscillation. The PD (Finisar, HPDV 2120R) is with the 3dB bandwidth of 40 GHz and responsivity of 0.8 A/W. Two EAs are cascaded to compensate the loss of the electrical signal. The bandpass filter (BPF) with 3 dB bandwidth of 2 GHz and central frequency of 5.5 GHz is employed to filter out the spurs and noise, and the filtered signal then is fed back to the RF port of the PM to form a closed loop.

The frequency response of proposed dual-band MPF is measured and presented in Fig. 2(a), which displays the dual frequency peaks at the 4.9 GHz and 6.2 GHz, respectively. Fig. 2(b) is the frequency spectrum of the generated continuous signal, and two single resonance frequency around 4.9 GHz and 6.2 GHz are demonstrated. And the signal to noise ratio (SNR) reaches 54 dB with clean background thanks to the matched electrical BPF. The corresponding temporal waveform is depicted in Fig. 2(c), and the inset is the enlarged picture, implying a high frequency oscillation modulated by a smaller frequency signal. The instantaneous frequency distribution of the generated waveform is calculated by short-time Fourier transform of the temporal waveform and is presented in Fig. 2(d). And it reveals dual constant oscillation single frequency around 4.9 GHz and 6.2 GHz, which is consistent with the frequency response of the MPF in Fig. 2(a). The state of the generation of two-single-frequency-signal is the named working mode 1.

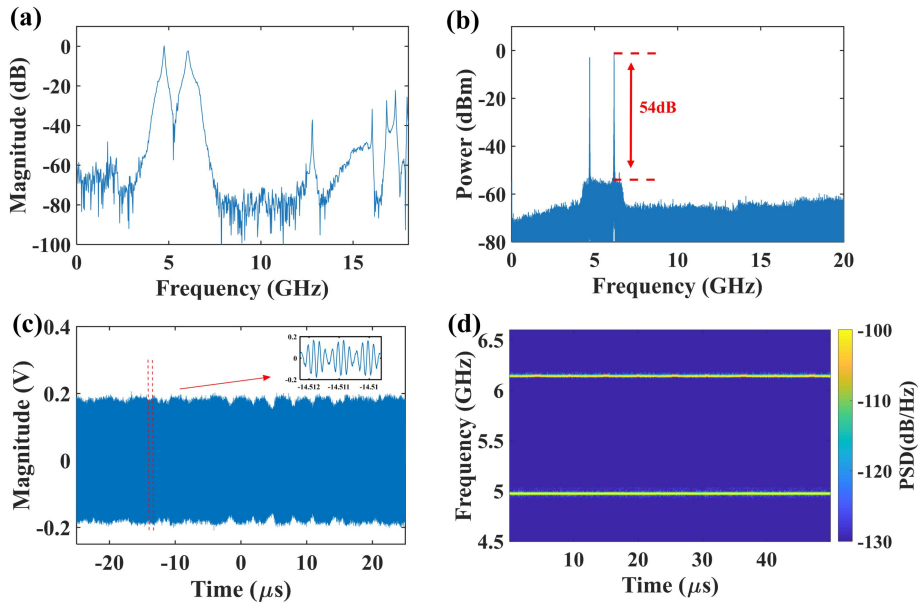

Figure 2 (a) The frequency response of proposed dual-band MPF; (b) The frequency spectrum of the generated dual-single-frequency signal; (c) Measured temporal waveform of the constructed OEO system at static state and the corresponding enlarged figure; (d) Corresponding instantaneous frequency distribution of the waveform of the generated dual-single-frequency signal

Then the spontaneous frequency-hopping is implemented by tuning the controller 3 (PCs) to regulate the power contribution, shown in Fig. 3. Fig. 3(a) is the frequency spectrum of the dual-frequency hopping signal, dual frequencies around 6 GHz and 8.8 GHz are demonstrated. The temporal waveform of the spontaneous frequency-hopping microwave signal is shown in Fig. 3(b). The temporal waveform indicates that the two groups of frequency-hopping modes have slightly different amplitudes due to the different power contribution in the loop, which can also be seen in the frequency spectrum in Fig. 3(a). The inset in Fig. 3(b) is an enlarge picture to show the details of the temporal waveform when frequency hopping occurs, indicating that the frequency-hopping speed reaches ~50 nanoseconds. The instantaneous frequency distribution of the generated frequency hopping waveform is calculated and displayed in Fig. 3(c). As can be seen from Fig. 3(c), the frequency sequence is periodic, and the period is equal to the cavity round-trip time of the OEO, which is about 5.12 μs in this case.

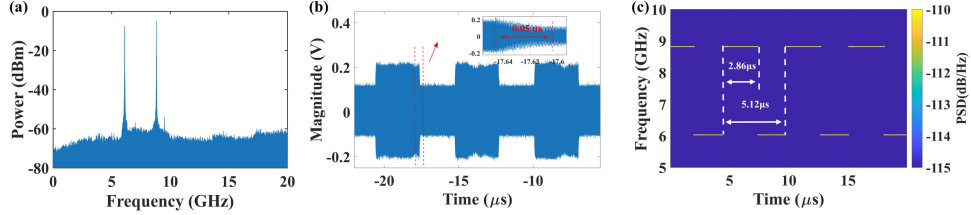

Figure 3 (a) The frequency spectrum of the dual-frequency hopping signal; (b) Measured temporal waveform of the generated dual-frequency hopping signal and the corresponding enlarged figure; (c) Corresponding instantaneous frequency distribution of the waveform of the generated dual-frequency hopping signal.

Since the power distribution of the pulse width of sequence is related to the cavity gain of the OEO loop (reference), pulse width tunning of the dual frequencies is discussed in Fig. 4(a), 4(b) and 4(c). By controlling the polarization state of the pump light signal and the light signal, the power distribution of the dual-frequency hopping signal is regulated, resulting in the changing of the pulse width of the sequence. Fig. 4(a), 4(b) and 4(c) demonstrate a changing of the pulse width ratio (PWR) of higher frequency to lower frequency changing from 1.56 μs to 3.38 μs. The central frequency of the dual- frequency hopping signal also can be easily tuned by changing the central frequencies of the LDs. Fig. 4(d), 4(e) and 4(f) illustrate the frequency tunability with fixed PWR. By changing the LDs, the frequency of the lower frequency signal is changing from 4.4 GHz to 7.9 GHz and the higher frequency is tuned from 6.2 GHz to 8.5 GHz. It should be noted that, the polarization state should be retuned at different central frequency due to the polarization frequency dependence.

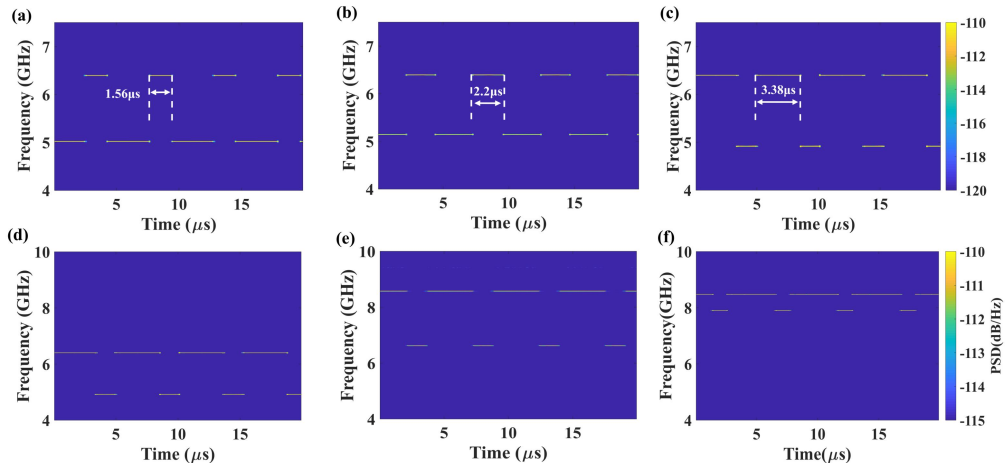

Figure 4 Corresponding instantaneous frequency distribution of the dual-frequency hopping signal. Difference pulse distribution of the dual frequencies is demonstrated in (a), (b) and (c); the frequency tunability with fixed PWR is demonstrated in (d), (e) and (f).

At the mode 2 working state, a driving signal with a period of 5.12 μs is applied to LD2 to form a frequency-scanning MPF based on the SBS effect. The FDML-OEO is achieved by synchronizing the period of the driving signal of the LD2 with the round-trip time of the OEO loop. Fig. 5(a) is the frequency spectrum of the generated broadband signal and single-frequency signal, which demonstrate two different kinds of MPF can work in the same OEO cavity. The corresponding temporal waveform is shown in Fig. 5(b). The waveform has nonuniform intensity fluctuation with a period of 5.12 μs which is consistent with the round-trip time of the OEO loop. The time width of the higher intensity partis 0.81 μs, the amplitude of which is a combination of the single frequency signal and the nonlinear broadband signal. The instantaneous frequency distribution of the generated signal is calculated and illustrated in Fig. 5(c). Clearly, the signal is composed by a pulsed single-frequency signal and a lower-band continuous nonlinear broadband frequency signal. This is the first time that RF signals with different modulation characteristics are generated in the same optoelectronic oscillator. This kind of signal can be used in multifunction radar to realize the large detection range and the high detection accuracy at the same time benefiting from the coherent single frequency pulse and broadband signal.

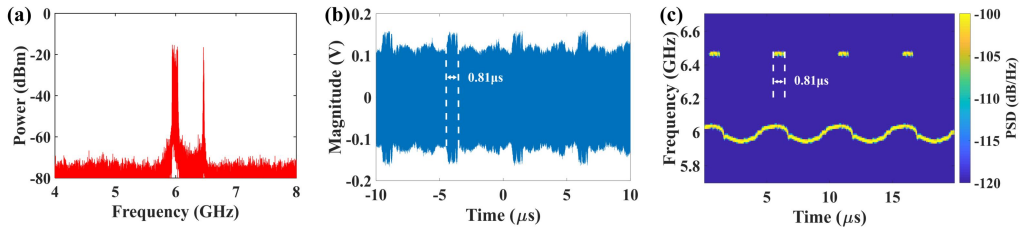

Figure 5 (a) The frequency spectrum of the generated signal with a broadband and a single frequency signal. (b) Measured temporal waveform of the generated broadband signal and the single frequency signal. (c) Corresponding instantaneous frequency distribution of the waveform of the generated broadband signal and the single frequency signal.

Fig. 6 is the flexibility illustration of this mode. The central frequency and bandwidth can be easily tuned by just changing the wavelength of LDs and the driving current of their controllers. Fig. 6(a) demonstrates the central frequency tunability of the dual-band signal. Fig. 6(b) illustrates that the bandwidth of the generated broadband signal is changing with the driving current voltage of the LD2. The bandwidth is changing from 50 MHz to 150 MHz by applying the driving current from 110 mVpp to 250 mVpp. Since the power contribution is related to the polarization, the single frequency pulse width can be tuned by controlling the polarization state of the output signals of the LDs. The time widths of the generated single-frequency signals in Fig. 5(c), 5(d), and 5(e) are 0.78 μs, 1.10μs, and 1.43 μs, respectively.

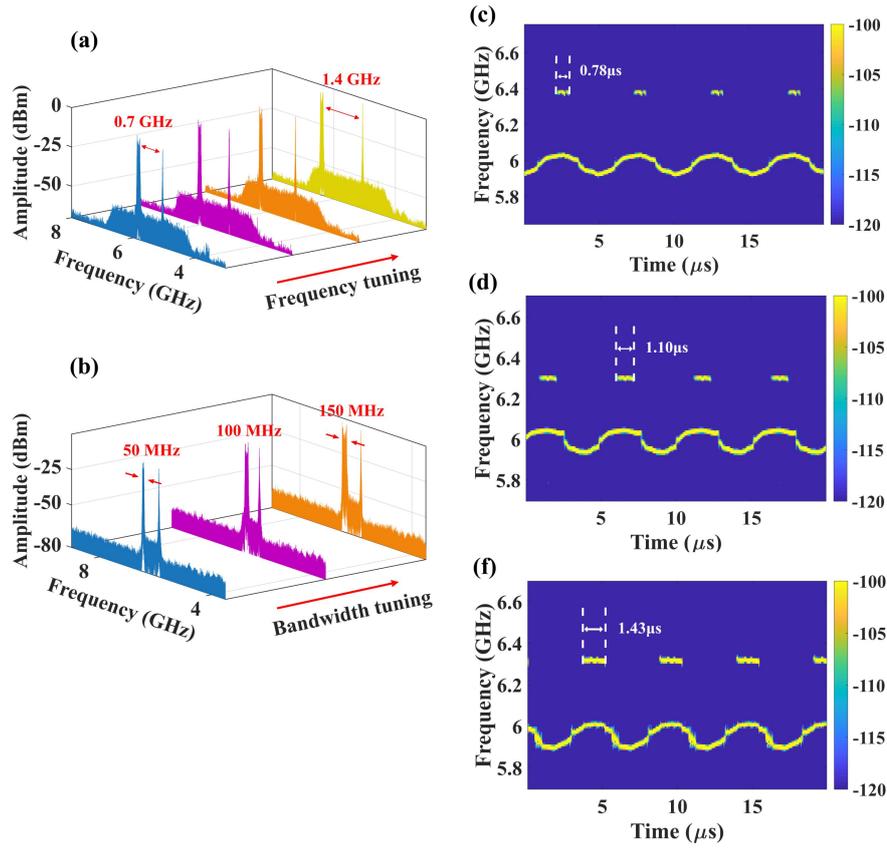

Figure 6 Experimental results of the tunability of generated signals at model 2. (a) The spectrum of the generated broadband signal under different central frequencies. (b) The spectrum of the generated broadband signal under different bandwidths. And the tunable time widths of the generated single frequency signals are demonstrated in (d), (e) and (f).

At the mode 3 working state, a driving signal with a period of 5.12 μs is applied to the LD1 to form a frequency-scanning MPF based on the PS-FBG. In this mode, the dual-band MPF are all working at broadband state, resulting in the generation of a dual broadband signal, shown in Fig. 7. Fig. 7(a) is the frequency spectrum of the generated signal, which displays a dual broadband signal at 4.7 GHz and 5.9 GHz with bandwidth of 600 MHz. The corresponding temporal waveform is depicted in Fig. 7(b). And Fig. 7(c) is the calculated instantaneous frequency distribution of the signal. In this case, the signal linearity and the oscillation state are more stable compared to that of mode 2, which is because of the instability of the sweeping SBS effect. The bandwidth and central frequency are also easily regulated by changing the LDs and the driving controller of LD1. The central frequency tunability of the generated dual-broadband signal is demonstrated in Fig. 7(d). The frequency of the lower frequency signal is changing from 4.7 GHz to 10.3 GHz and the higher frequency is tuned from 6.2 GHz to 12.3 GHz. The corresponding bandwidth tunability is shown in Fig. 7(e), in which the bandwidth of the generated dual-broadband signal is tuned from 250 MHz to 700 MHz by changing the driving current volage of LD1.

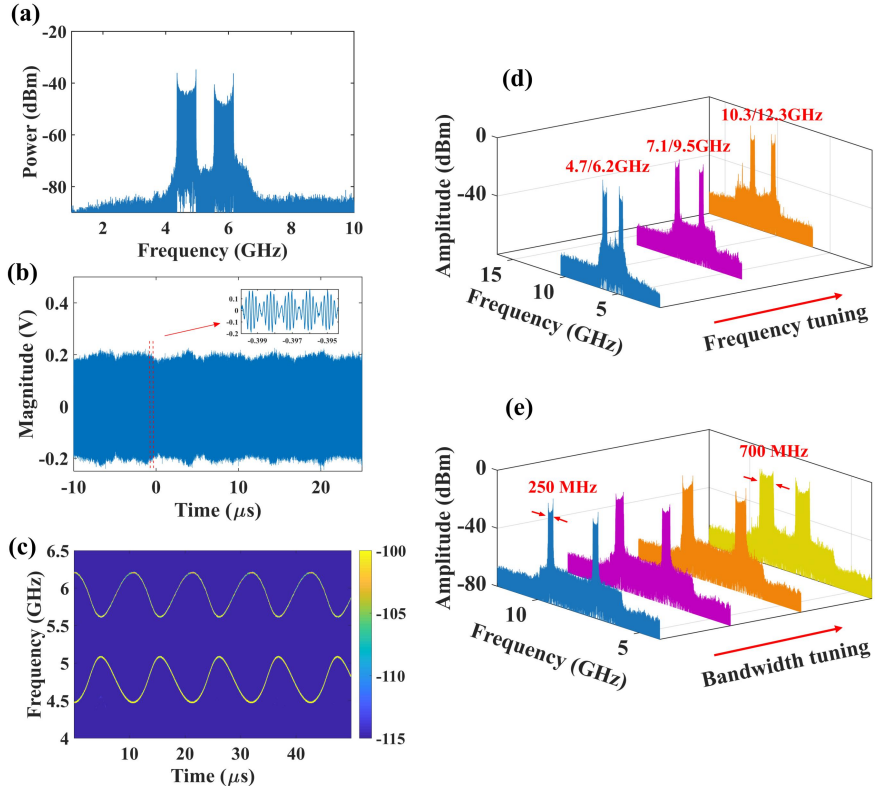

Figure 7 (a) The frequency spectrum of the generated dual-broadband signal; (b) Measured temporal waveform of the generated dual-broadband signal and the corresponding enlarged figure; (c) Corresponding instantaneous frequency distribution of the generated waveform; (d) The spectrum of the generated dual broadband signal under different central frequencies. (b) The spectrum of the generated dual broadband signal under different bandwidths.

The phase noise and detection performance of the proposed OEO is also evaluated and shown in Fig. 8. The single sideband (SSB) phase noise of the OEO working at dual-band MPF based on SBS and FBG respectively is measured and given in Fig. 8(a). The phase noise of generated single frequency signal based on SBS effect is −89 dBc/Hz at 10 kHz offset while the phase noise based on FBG is -112 dBc/Hz at 10 kHz offset with the fiber length is ~1 km. With longer fiber or stable environmental state, the data will be also improved. To evaluate the detection performance of the generated broadband signal, pulse compression of the signal in Fig. 7(c) is carried out, result of which is shown in Fig. 8(b). It is obvious that the distance between adjacent peaks in the Fig. 8(b) is the same as the period of the generated broadband signal shown in Fig. 7(c). And the inset is the enlarged picture, implying a compressed time width of the ~1ns, which indicates a well coherent of the generated dual wideband signals.

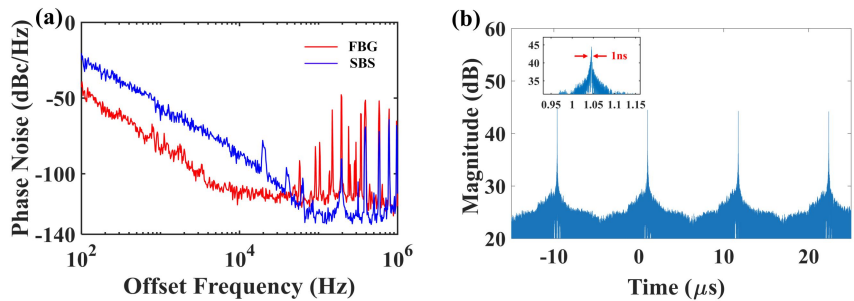

Figure 8 (a) Measured phase noise performance of the dual single frequency signal based on SBS (blue line) and

FBG (red line) respectively; (b) The result of pulse compression of the generated dual broadband signal and the corresponding enlarged figure.

**Conclusion**

In conclusion, a method with reconfigurable multiple formats for simultaneous generation of coherent dual band signal is proposed and experimentally demonstrated. Three different working states of the proposed scheme are demonstrated and reconfigured by regulating the external controllers without any changing of the system. At mode 1 state, a dual single frequency hopping signal is achieved with 50 ns hopping speed, flexible central frequency and tunable pulse duration ratio. The mode 2 state is realized by applying the Fourier domain mode-locked technology into the proposed optoelectrical oscillator, in which a dual coherent signal with a pulsed single frequency signal and a broadband signal is achieved. The duration of the single frequency signal and the bandwidth adjustability of the broadband signal are discussed. When the system working at the mode 3 state, a dual broadband signal generator which is realized by injecting a triangular wave into the signal laser. The detection performance of the generated broadband signals has also been evaluated by the pulse compression and the phase noise figure. The proposed method may provide a multifunction radar signal generator based on the simply external controllers, which can serve in a low-phase-noise radar system and multifunction of an early warning detection radar system with high resolution imaging in complex interference environment.

spontaneous frequency-hopping optoelectronic oscillators," Photon. Res. **10**(5), 1280-1289 (2022)